# Bayesian Analysis for Stellar Evolution with Nine Parameters (BASE-9) v9.4.3

## User's Manual

Ted von Hippel[1], Elliot Robinson[2], Elizabeth Jeffery[3], Rachel Wagner-Kaiser[4], Steven DeGennaro[5], Nathan Stein[6], David Stenning[7], William H Jefferys[8], and David van Dyk[9]

[1]Embry-Riddle Aeronautical University, Daytona Beach, FL, USA; ted.vonhippel@erau.edu
[2]Argiope Technical Solutions, Ft White, FL, USA; elliot.robinson@argiopetech.com
[3]Brigham Young University, Provo, UT, USA; ejeffery@byu.edu
[4]University of Florida, Gainesville, FL, USA; rawagnerkaiser@gmail.com
[5]Studio 42, Austin, TX, USA; studiofortytwo@yahoo.com
[6]University of Pennsylvania, Philadelphia, PA, USA; nathanmstein@gmail.com
[7]University of California, Irvine, CA, USA; dstennin@uci.edu
[8]University of Texas, Austin, TX, USA and University of Vermont, Burlington, VT, USA; bill@astro.as.utexas.edu
[9]Imperial College London, London, UK; d.van-dyk@imperial.ac.uk

BASE-9 is a Bayesian software suite that recovers star cluster and stellar parameters from photometry. BASE-9 is useful for analyzing single-age, single-metallicity star clusters, binaries, or single stars, and for simulating such systems. BASE-9 uses Markov chain Monte Carlo and brute-force numerical integration techniques to estimate the posterior probability distributions for the age, metallicity, helium abundance, distance modulus, and line-of-sight absorption for a cluster, and the mass, binary mass ratio, and cluster membership probability for every stellar object. BASE-9 is provided as open source code on a version-controlled web server. The executables are also available as Amazon Elastic Compute Cloud images. This manual provides potential users with an overview of BASE-9, including instructions for installation and use.

# Table of Contents

# I. Introduction

Bayesian Analysis for Stellar Evolution with Nine Parameters (BASE-9) is a Bayesian software suite that recovers star cluster and stellar parameters from photometry. BASE-9 is useful for analyzing single-age, single-metallicity star clusters, binaries, or single stars, and for simulating such systems. This document assumes you are working with base 9.4.3. We will endeavor to update this manual as we update the code or as libraries or operating systems meaningfully change.

BASE-9 uses a Markov chain Monte Carlo (MCMC) technique along with brute-force numerical integration to estimate the posterior probability distribution for up to six cluster and three stellar properties. The cluster properties are age, metallicity, helium abundance, distance modulus, line-of-sight absorption, and parameters of the initial-final mass relation (IFMR). The stellar properties are primary mass, secondary mass (if a binary), and cluster membership probability. The MCMC technique is used for the cluster quantities and numerical integration is used for the stellar quantities. BASE-9 is freely available source code that you may use as is or modify for your own research and educational purposes.

BASE-9 may be the code for you if
1. you are dissatisfied with deriving cluster-level parameters by over-plotting isochrones on your data and iteratively adjusting parameters,
2. you wish to recover more than just an average and error bar for each parameter, and instead wish to characterize the probability distributions for these parameters,
3. you wish to take fuller advantage of ancillary data, such as proper motion membership probabilities, spectroscopic mass estimates, or distances from trigonometric parallaxes.

This manual is designed to help you install and run BASE-9. If you use BASE-9 in your research, please cite

    von Hippel, T., Jefferys, W. H., Scott, J., Stein, N., Winget, D. E., DeGennaro, S., Dam, A., & Jeffery, E. 2006, *Inverting Color-Magnitude Diagrams to Access Precise Star Cluster Parameters: A Bayesian Approach*, ApJ, 645, 1436

and if you find the following helpful, please also cite

    DeGennaro, S., von Hippel, T., Jefferys, W. H., Stein, N., van Dyk, D. A., & Jeffery, E. 2009, *Inverting Color-Magnitude Diagrams to Access Precise Star Cluster Parameters: A New White Dwarf Age for the Hyades*, ApJ, 696, 12

    van Dyk, D. A., DeGennaro, S., Stein, N., Jefferys, W. H., & von Hippel, T. 2009, *Statistical Analysis of Stellar Evolution*, Annals of Applied Statistics, 3, 117

Depending on how you use BASE-9 (this part is under your control), the software also relies on the stellar evolution models of

    Dotter, A., Chaboyer, B., Jevremovic, D., Kostov, V., Baron, E., & Ferguson, J. W. 2008, *The Dartmouth Stellar Evolution Database*, ApJS, 178, 89

Williams, K. A., Bolte, M., & Koester, D. 2009, *Probing the Lower Mass Limit for Supernova Progenitors and the High-Mass End of the Initial-Final Mass Relation from White Dwarfs in the Open Cluster M35 (NGC 2168)*, ApJ, 693, 355

or a fitted IFMR parameterized as lines, broken lines, or low-order polynomials as described by

Stein, N. M., van Dyk, D. A., von Hippel, T., DeGennaro, S., Jeffery, E. J., & Jefferys, W. H. 2013, *Combining Computer Models in a Principled Bayesian Analysis: From Normal Stars to White Dwarf Cinders*, Statistical Analysis and Data Mining, 6, 34

For a further discussion of what BASE-9 and its precursor, BASE-8, has been used for to date and some indications of how it might be useful in your research, see also the following papers:

Jeffery, E. J., von Hippel, T., Jefferys, W. H., Winget, D. E., Stein, N., & DeGennaro, S., 2007, *New Techniques to Determine Ages of Open Clusters Using White Dwarfs*, ApJ, 658, 391

Jeffery, E. J., von Hippel, T., DeGennaro, S., Stein, N, Jefferys, W.H., & van Dyk, D. 2011, *The White Dwarf Age of NGC 2477*, ApJ, 730, 35

## II. Skip the install and go to the cloud

BASE-9 executables are available as Amazon Elastic Compute Cloud (EC2) images. Up-to-date instance IDs are listed in the release descriptions at http://github.com/argiopetech/base/releases. An Amazon Web Services (AWS) account is required to use these instances.

To run your code on EC2,
1. Log in to your AWS account
2. Navigate to EC2
3. Navigate to the "Instances" pane
4. Click "Launch Instance"
5. Choose "Community AMIs"
6. Enter the AMI code for the version of BASE-9 you would like to run
7. Select "Review and Launch", then "Launch"
8. Wait for the instance to launch
9. use rsync/scp to copy your data to the public IP of the instance
10. Login to your instance with SSH

The default user name is `ec2-user`. There is no root password by default.

BASE-9 executables are in `/usr/local/bin` (should be in the path). Current models (appropriate for the installed version of BASE-9) are in `/usr/local/share/base-models`.

The instance operating system is the newest release of FreeBSD 10. The tcsh, csh, sh, and bash shells are available. VI, VIM, Emacs, and nano are pre-installed.

## III. Installation

BASE-9 is written in C++ and designed to run on a variety of UNIX- and Linux-based operating systems. It is currently tested on

- Mac OS X 10.7 through 10.9
- Ubuntu 10.04 through 12.04
- RHEL 5 and 6
- Gentoo 13.0
- FreeBSD 9 and 10

To compile the code you will need gcc 4.7+ or clang 3.2+ (C/C++ language compilers), gsl (the gnu science library), cmake (a cross-platform build system), and Boost (a peer-reviewed, portable C++ library). To install these software packages, you may need help from your system administrator, though we provide some guidance here.

The best place to put all of this code is in the /usr/local/bin directory. If you don't have that directory on your machine already, you can create it as follows:

```
> sudo mkdir /usr/local
> sudo mkdir /usr/local/bin
```

Note that the sudo command gives you super-user or root permission for that one command (after you enter your password at the prompt), assuming that your account has been allowed to invoke the command.

### A. Installing gcc, gsl, and cmake on a Mac running OS X 10.7 – 10.9

Download the compiler. One way to do that is via downloading Xcode 4.6 or later from http://connect.apple.com. This requires that you have a developer account, but you can register for that for free. Also, it will give you 1+ GB of code and tools, most of which you will only need if you intend to develop for iPhones, Mac OS, etc. If you do follow this route, after installing Xcode, you will need to specifically install the command line tools with a window that will look similar to the one below. Click on the install button to the right of "command line tools" and it will appear as follows when done.

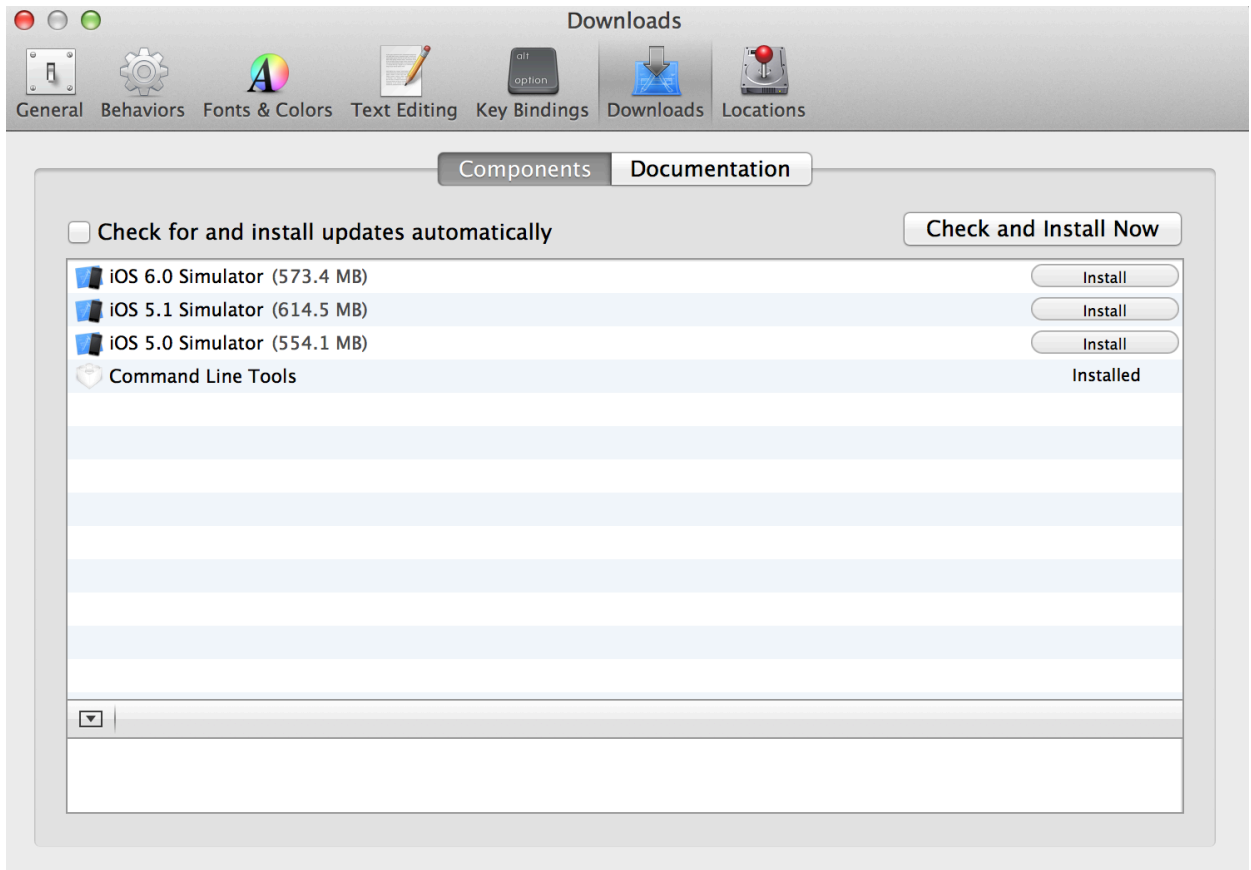

This will install clang and (for Xcode 4.6) GCC 4.2.

Download gsl from http://www.gnu.org/software/gsl/. Use the ftp site to obtain the source code, then

```
mac> cd ~/Downloads
mac> tar xzf gsl-1.15.tar.gz
mac> cd gsl-1.15/
mac>./configure
mac> make
mac> sudo make install
```

Download cmake from http://www.cmake.org/cmake/resources/software.html. Choose the .dmg version of the code (for the correct operating system) if you want to let Mac installation guide you through the process. We suggest placing the cmake build into `/usr/local/bin` by choosing that directory when prompted.

## B. Installing gcc, gsl, and cmake on a Linux machine

The simplest way to install on Ubuntu is via the apt-get tool.

```
linux> sudo apt-get install gcc
linux> sudo apt-get install cmake cmake-curses-gui
linux> sudo apt-get install libgsl0-dev libgsl0ldbl
```

A similar process will work with the yum tool on Fedora / RHEL* :

```
linux> sudo yum install gcc gcc-c++ cmake git
linux> sudo yum install gsl gsl-devel boost boost-devel
```

*RHEL 5 & 6 repositories have an old gcc version. The devtoolset package will install an alternate, up to date build environment at /opt/<distro>/devtoolset-2/

```
linux> sudo yum install devtoolset-2-toolchain
```

The scl utility can create a shell referencing these alternate build tools where BASE-9 can be built:

```
linux> scl enable  devtoolset-2 'bash'
```

## C. Unpacking BASE-9

Create a directory where you wish to install and run the software, then download the newest code release from https://github.com/argiopetech/base/tags and the newest stellar evolution files from https://github.com/argiopetech/base-models/tags, and extract them to the appropriate directory, e.g.

```
> tar xzf base-9.4.2.tar.gz
> cd base-9.4.2/
```

Note that your computer may uncompress the .gz file for you on download, in which case the above command would instead be

```
> tar xvf base-9.4.2.tar
> cd base-9.4.2/
```

### D. Installing Boost

BASE-9 has an included script to install Boost. The install location can be changed by modifying the `CMAKE_INSTALL_PREFIX` variable.

```
linux> cd contrib
linux> cmake –DCMAKE_INSTALL_PREFIX="/usr/local" .
linux> sudo make
```

Ubuntu users can save some time on this step by running

```
linux> sudo apt-get install libboost-dev
```

### E. Installing BASE-9

Once you have all of the above software in place, you are ready to install BASE-9. The following instructions should work identically for all platforms.

Change directories into the BASE-9 source directory and simply run build.sh

```
> sudo ./build.sh
```

This will (if you have properly installed all libraries) build and install the BASE-9 executables and install them in the default location (generally `/usr/local/bin`)

Alternatively, if you do not have the ability to run `sudo` on your machine, you may use

```
> ./build_local.sh
```

to build and install the executables locally. The executables will be installed in the `BUILD/bin` directory.

### IV. Running BASE-9

In the following subsections, we describe how to run stand-alone portions of the BASE-9 modules from the command line. There are various reasons why you might want to run one or another of these, or some, but not all, so we detail how to run each one. As of BASE-9.2.0, all settings have been moved into a YAML-format configuration file. A sample configuration file with reasonable initial settings can be found in the `base-9.4.2/conf` directory under the name `base9.yaml`. A sample cshell script can be found at `scripts/hyades.csh`. Individual settings can be changed on a run-by-run basis via the command line options. Run any of the BASE-9 applications with the command line flag "--help" to view a description of available settings.

The following examples assume you have installed the BASE-9 executables in a directory that is in your `PATH` (e.g., `/usr/local/bin`). If this is not the case, you may need to use absolute pathnames (e.g., `/home/me/base-9.4.2/BUILD/bin/singlePopMcmc`).

## A. simCluster

The first tool that you are likely to use within BASE-9 is `simCluster`. This module simulates a stellar cluster for a particular set of models (see references in the Introduction) and user-specified values of various cluster parameters that have been set in the base9.yaml file: the base-10 log of the cluster age, metallicity, the helium mass fraction (only for some Dotter et al. models), the distance modulus, absorption in the V-band, the percent of cluster stars that are binaries, the upper mass limit for creating a white dwarf (WD), and the fraction of WDs that have helium atmospheres (DBs). We recommend leaving these last two parameters at 8 solar masses and 0%, as we have not yet fully implemented and tested them. An additional parameter, the seed to the random number generator, is necessary as the mass of each star is determined by randomly drawing from the IMF. This allows you to specify multiple clusters with the same parameters but different random seeds if you wish to test the effects of, for instance, cluster size on the number of WDs or the clarity of the main sequence turn-off (MSTO). This seed can be set via the `--seed` option in the command line. To run `simCluster`, simply type its name:

```
linux> ./simCluster
Seed: 1559729633
Reading models... Done.

Properties for cluster:
 logClusAge     =  8.796
 [Fe/H]         =  0.07
 Y              =  0.29
 modulus        =  0.00
 Av             =  0.01
 WDMassUp       =  8.0
 fractionBinary =  0.00
Totals:
 nSystems       = 100
 nStars         = 100
 nMSRG          = 98
 nWD            = 2
 nNSBH          = 0
 massTotal      = 62.72
 MSRGMassTotal  = 55.10
 wdMassTotal    = 1.66
```

The above output is diagnostic and reiterates the settings in the `base9.yaml` file. The stored output of `simCluster` is placed in a filename specified by the user with the `outputFileBase` option in the `base9.yaml` file. The file contents from the output of `simCluster` should look like the following:

```
linux>head -2 hyades.sim.out
   id     U       B       V       R       I       J       H       K      sigU
sigB    sigV    sigR    sigI    sigJ    sigH    sigK   mass1 massRatio stage
Cmprior useDBI
    1  3.061   3.031   2.694   2.501   2.320   2.127   1.983   1.965   0.000
0.000   0.000   0.000   0.000   0.000   0.000   0.000   1.555     0.000
1   0.999       1
```

There are more columns than can be presented cleanly on a page, but hopefully this is clear enough. The first column lists identification numbers for each star system (single star or binary). This is meant to be useful in tracking down particular stars. The next eight columns list the U-through K-band magnitudes (or ugriz through K) of the primary star. Columns 10 through 17 give the photometric uncertainties for each filter entry (for a simulated cluster, these are zero).

The 18$^{th}$ and 19$^{th}$ columns give the mass of the primary star and the mass of its companion (if applicable). The 20$^{th}$ column lists the stage of stellar evolution for that particular star (1 = MS or RG, 3 = WD, >3 for evolved stars above the WD mass limit). The final two columns are the cluster membership prior (which is essentially ~1 for simulated stars) and the flag (0 or 1) whether to use the star during the burn-in stage. With these final columns, the output file is formatted for input into `scatterCluster`. You should be able to plot reasonable looking CMDs/isochrones from this file for a wide range of cluster parameters, stellar models, and filters.

### B. scatterCluster

The `scatterCluster` module adds Gaussian random errors to the photometry output created by `simCluster`. To specify the appropriate amount of error to add for your particular simulation, adjust the virtual exposure time in the `base9.yaml` file.

We use exposure times of 1 hour in each filter to generate a scattered cluster with the above file. The algorithm for adding noise to the cluster photometry is rudimentary and only meant for simple purposes such as preparing for an observing proposal or for creating test files for the Markov chain Monte Carlos (`singlePopMcmc`) routine. The algorithm is an approximation to the results one would obtain in one hour with the KPNO 4m + Mosaic (UBVRI) or Flamingos (JHK), assuming dark time, seeing=1.1 arcsec, airmass=1.2. Signal-to-noise for the Spitzer bands, if included, are naively set to be the same as for the K-band. For departures from a one-hour exposure the S/N is scaled by sqrt(exptime). These exposure times can be set in the `base9.yaml` file under `exposures` for each individual filter.

Additional options for `scatterCluster` are available in the yaml file. The bright and faint end cut-off mags allow you to narrow the portion of the CMD that you wish to retain. The `relevantFilt` option specifies which band is the reference filter (in this case, 0=U, 1=B, etc.). The `base9.yaml` options `brightLimit` and `faintLimit` refer to the bright and faint end cut-off magnitudes for the reference filter indicated. You can also clip on S/N with `limitS2N` and decide to cut out field stars, if they were simulated by `simCluster`. Additionally, `scatterCluster` will determine which filters you are using based on the header in the `simCluster` output file. Again, the integer `seed` may be set at the command line to allow you to start from the same input file, but create multiple simulated observations of that file with different initial `seed` values.

```
linux> scatterCluster
Seed: 1564704505
```

The output file of `scatterCluster` looks like

```
linux> head -2 hyades/hyades.scatter.out
  id      U      B      V      R      I      J      H      K    sigU
sigB   sigV   sigR   sigI   sigJ   sigH   sigK    mass1 massRatio stage
Cmprior useDBI
   1  3.065  3.016  2.691  2.509  2.328  2.128  1.985  1.977   0.010
0.010  0.010  0.010  0.010  0.010  0.010  0.010    1.555         0.000
1  0.999        1
```

Notice now that the output includes the estimated errors for each band (`sig-`). The format of the output file is otherwise the same as the input file for `scatterCluster`.

In this case, only the `id`, `mass1`, and `stage1` values are kept from the output of `simCluster`. The photometry values (here UBVRIJHK) are derived from the photometry values in the `simCluster` output file, but are different in that they are scattered by adding a Gaussian random deviate with sigma = `sigU`, `sigB`, etc. This section of the output file is all one needs to plot realistic CMDs for proposals and possibly to prepare for observing projects. The `scatterCluster` output file contains additional information, however, and is formatted to be ingested by `singlePopMcmc`, so that it can be used to test `singlePopMcmc` and so that you can test the precision and accuracy that you would expect to recover from real data based on a given set of cluster parameters, observational errors, and the number of stars available. The `massRatio` column lists the ratio by mass of the secondary to primary stars, which in these examples are both 0 since there were no secondaries. The `CMprior` column is set by default in `scatterCluster` to 0.99, but the file can easily be edited to set a different prior probability that any particular star is a cluster member. The final column is just a 0 or 1 switch (off or on) of whether to use a particular star during the burnin process. (See DeGennaro et al. 2009 and van Dyk et al. 2009 for a discussion of what the burnin entails and why it is used.) To make it easiest for `singlePopMcmc` to converge, it is helpful to have this parameter set to 1 for stars that are likely to be cluster members and if there are many field stars, it is helpful if the bulk of them can be set to 0 at this point.

# C. singlePopMcmc

The `singlePopMcmc` module is the workhorse of our software suite. This routine, along with its many subroutines, runs a Markov chain Monte Carlo sampler using a variety of standard Bayesian techniques as well as a few techniques newly developed by us. The approach and mathematics are presented by DeGennaro et al. (2009), van Dyk et al. (2009), and Stein et al. (2013). This code was designed to run on photometry formatted in the same manner as the output of `scatterCluster`. It can also be run just as easily on the simulated photometry from `simCluster` + `scatterCluster`.

The `singlePopMcmc` module has a variety of values and options set in the `base9.yaml` file. Under the `singlePopMcmc` group, the `stage2IterMax` and `stage3Iter` set the length of the burnin for `singlePopMcmc`. The `runIter` option lets you choose the number of iterations of the Markov chain Monte Carlo. The rule-of-thumb is that one typically wants 10,000 well-sampled points from a Markov chain Monte Carlo in order to draw robust inferences on the posterior distribution. At the other extreme, the Central Limit Theorem dictates that approximately 30 uncorrelated samples are sufficient for a normal distribution. Before running a particular dataset against a specific set of models, you do not know if the posterior distributions will be Gaussian shaped or more complex, so we suggest you take the conservative approach and initially assume complex posterior distributions and run BASE-9 for 10,000 uncorrelated iterations. The parameter `thin` sets the increment between saved iterations. We recommend that this parameter be left equal to 1 to keep the adaptive sampling routine efficient. If the output of `singlePopMcmc` is correlated (see below), then each new iteration or step is not independent and you need substantially more than 10,000 iterations to draw robust inferences. In situations like this, we recommend that the output file be thinned afterwards, i.e. that the user uses every n$^{th}$ record where n is large enough to keep the output uncorrelated.

Under the `cluster` options, there are five parameters for which means and standard deviations can be set: the metallicity prior (`Fe_H`), the distance modulus prior (`distMod`), the absorption prior (`Av`), the helium prior (`Y`), and the carbon fraction prior for a C+O WD (`carbonicity`). Note that `carbonicity` only works with the Montgomery models and is not yet supported because we are currently testing it. If you only have weak priors, that is fine. If you do not want to sample on one or more of these parameters, you can set the `sigma` for that parameter to 0.0 and this will turn of sampling for that parameter. Under `starting`, the parameter `logClusAge` is a starting value for the log of the age in years (e.g. 9.0 for a 1 billion year old cluster). *This is not a prior*, but just tells `singlePopMcmc` where to start searching for a fit. We have found that although convergence may depend on starting with a roughly reasonable age, the actual posterior age distribution does not depend on what that value is, assuming it does converge.

The `msRgbModel` lets you choose which set of models to use with your data (the filters available in the models must match the filters of your observed or simulated/scattered cluster). This allows you to derive cluster parameters for a range of models as well as to create simulated clusters under one set of models and use `singlePopMcmc` to derive the cluster and stellar parameters under another set of models. The latter experiments might be useful, for instance, if you wanted to test the sensitivity of basic cluster or stellar parameters to a given model ingredient. With ancillary data for cluster or stellar parameters this might allow you to constrain model ingredients.

Again we mention that the seed can be set inline with `--seed` when `singlePopMcmc` is called. If `singlePopMcmc` appears to be unable to converge on reasonable cluster values, rerun it with a different initial seed. Changing the seed also allows you to start a new MCMC chain if you ran a prior calculation with too few iterations.

To run `singlePopMcmc`, using a properly prepared input `base9.yaml` file, type the following:

```
linux> singlePopMcmc --verbose
Bayesian Analysis of Stellar Evolution
Seed: 1570065938
Reading models... Done.

Model boundaries are (7.800, 10.250) log years.
Binaries are OFF

Running Stage 1 burnin... Complete (acceptanceRatio = 0.090)

Running Stage 2 (adaptive) burnin...
    Acceptance ratio: 0.350. Trying for trend.
    Acceptance ratio: 0.600. Retrying.
    Acceptance ratio: 0.380. Trying for trend.
    Acceptance ratio: 0.520. Retrying.
    Acceptance ratio: 0.280. Trying for trend.
    Acceptance ratio: 0.180. Retrying.
    Acceptance ratio: 0.400. Trying for trend.
    Acceptance ratio: 0.440. Retrying.
    Acceptance ratio: 0.320. Trying for trend.
    Acceptance ratio: 0.500. Retrying.
    Acceptance ratio: 0.240. Trying for trend.
  Leaving adaptive burnin early with an acceptance ratio of 0.220 (iteration 1300)

Starting adaptive run...  Preliminary acceptanceRatio = 0.300
```

The `singlePopMcmc` routine creates multiple output files. In this case, it created:

```
-rw-r--r--   1 comp   staff    57955 Nov   3 17:27 hyades/hyades.res
-rw-r--r--   1 comp   staff    50317 Nov   3 17:25 hyades/hyades.res.burnin
```

The `.burnin` files provide the sampling patterns during the burnin process and may be useful for diagnostic purposes, especially if `singlePopMcmc` is not sampling well (see below). The `.res.burnin` files look like:

```
linux> head -2 hyades/hyades.res.burnin
     logAge            Y          FeH     modulus  absorption       logPost
   8.821886     0.280626     0.086646   -0.010790    0.011206   -174.520334
```

And the `.res` files have the same format:

```
linux> head -2 hyades/hyades.res
     logAge            Y          FeH     modulus  absorption       logPost
   8.843553     0.288468     0.016795   -0.180626    0.013647   -198.263339
```

After the column headers, there is one record for each iteration of each of the cluster parameters of interest.

If everything goes well, all you really need to do is plot histograms for any column of interest. These are the posterior parameter distributions. You can also calculate moments of these columns if you'd like, and look at correlations among the columns, e.g. by plotting logAge vs. modulus.

### D. sampleMass and sampleWDMass

These modules are useful for anyone interested in the masses of some or all of the stars in their database. Running them is unnecessary if you are only interested in the cluster parameters. The module `sampleMass` reports the primary mass and secondary mass ratio at all iterations for every star in the database, and `sampleWDMass` reports the primary mass for the subset of database stars that are being fit as WDs.

Running these programs is quite simple:

```
linux> sampleWDMass
Seed: 1690745648
Warming up generator… Done.
Generated 10000 values.
Reading models… Done.

sampledPars.at(0).age    = 8.78411
sampledPars.at(last).age = 8.74765
Part 2 completed successfully
```

Running `sampleMass` is effectively identical.

These output files names end with `.wdMassSamples`, `.wdMassSamples.membership`, `.massSamples`, and `.massSamples.membership`. These correspond to the WD mass outputs from `sampleWDMass`, the membership likelihood of those masses, the mass and secondary mass ratio outputs from `sampleMass`, and the membership likelihood of those pairs.

`sampleWDMass` output files consist of the same number of columns as there are WDs, and the same number of rows as there are in the results (`.res`) file. Each item in a row corresponds to the mass of a WD (ordered as in the database) given the sampled parameters in the results file. The membership file shares this format, though the values correspond to the likelihood that the given star is a member of a cluster with the given parameters.

`sampleMass` output files are similar to `sampleWDMass` output files but have two columns per star in the database. For every 0-indexed star $k$ in the database, column $2k$ corresponds to that star's primary mass, and $2k+1$ to that star's secondary mass ratio. The membership file is identical to that of `sampleWDMass`, though the values correspond to the likelihood that the given unresolved binary is a cluster member.

`sampleWDMass` has no configurable parameters.

`sampleMass` takes two parameters in the YAML file: `deltaMass` and `deltaMassRatio`. These values are used as starting step sizes for the adaptive MCMC process used to obtain mass and mass ratio. We recommended that you change these parameters only if you are manipulating the code for diagnostic purposes.

### E. makeCMD

The final module of our software suite, `makeCMD`, is a small module that calculates a mean fit isochrone. This is helpful for runs that do not converge as well as for situations where the posterior distribution of some key parameter may be multimodal. To run `makeCMD`

```
linux> makeCMD
Seed: 1574116425
Reading models... Done.

***Warning: "F435W" is not available in the selected WD Atmosphere
model
          This is non-fatal if you aren't using the WD models
```

The output of `makeCMD` looks like

```
linux> head -2 hyades/hyades.cmd
      Mass             U            B            V            R            I
J           H            K            F435W        F475W        F550M        F555W
F606W       F625W        F775W        F814W
   0.150000   16.170454   14.623905   13.035403   11.950929   10.490719
 9.196773    8.640356    8.386555   14.666132   13.907510   12.768529
13.128038   12.587469   12.243942   10.744691   10.477001
```

Because `makeCMD` uses the values of means under `cluster` in the `base9.yaml`, one can enter the mean or median values from the `singlePopMcmc` posterior distributions into the yaml file prior to running `makeCMD`. The output file from `makeCMD` can then be used to overplot what essentially amounts to the average fit isochrone from among the posterior parameter distributions. Note that this is not a best-fit isochrone, but rather a representative example drawn from that distribution. In fact, isochrones created from summary statistics such as mean or median parameters may not be truly representative if the distributions are substantially non-Gaussian because that simultaneous combination of parameters may fit the data with low probability.

### F. Hyades Test

We have created a script, `hyades.csh`, which is set up to run on a Hyades data set (`Hyades.UBV.testphot`). It is a cshell script. If you have problems with this script, you may be using a shell other than the cshell or tshell, e.g. the Bourne shell. You can invoke the cshell as follows:

```
Bourne shell> csh
New prompt indicating you are now running csh> hyades.csh
```

This will allow you to test your code installation and plot results, then compare to the DeGennaro et al. results. Note that you will not obtain an exact correspondence to the results of DeGennaro et al. because we have updated the Hyades data set since that publication. Because of the relative depth of the Hyades, which is significant compared to its distance, we have now corrected the cluster stars to lie at the mean cluster distance using individual proper motions from Hipparcos and the cluster converging point method. Because of the way we have corrected distances, this data set is converted to absolute magnitude space (we otherwise always use apparent magnitudes) and for this one test case, you will find a distance modulus of approximately 0.0.

### G. How long does all of this take?

In our tests, it took 147 minutes to run `hyades.csh`, which in turn ran `singlePopMcmc` for 152 Hyades stars in three photometric bands for 10,000 iterations on a early 2011 Macbook Pro (2.3 GHz Intel with 8 GB RAM) laptop computer. Increasing the number of filters or number of stars will increase the computation time linearly. Increasing the number of MCMC iterations will increase the run time, but somewhat less than linearly because some of the time is spent during the burnin. You will see substantial increases in runtime if you have much larger data sets and/or if you have to increase the total number of calculated iterations.

## V. Diagnostics of run quality

The following two plots show examples of poor and good sampling. In the first (extreme) case, the age sampling is highly correlated and one would need to use post-run thinning, probably by a factor of ~100. This means that one would need to run the code for 100x as many interations. The metallicity sampling displays only minor correlation, and if all other parameters looked this uncorrelated then this run would be sufficient. In this particular case, both plots were generated from the same `singlePopMcmc` run and because no single parameter is reliable until all parameters are essentially uncorrelated, this run did not reliably determine the metallicity (or any other) posterior distribution.

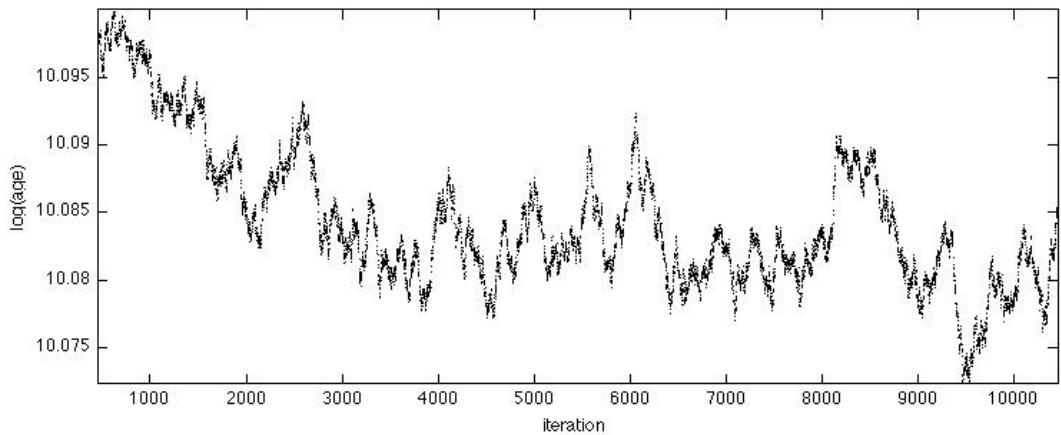

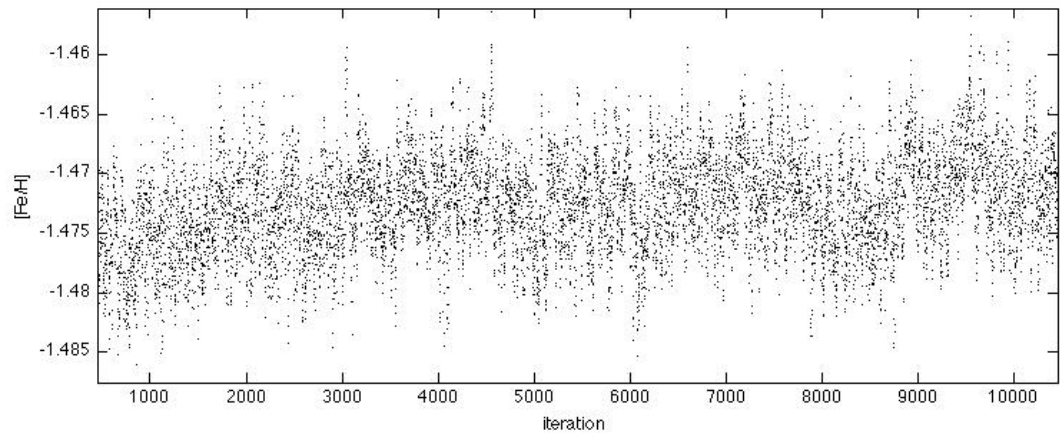

# VI. Example uses of BASE-9

In Section IV we outlined how to use the outputs of BASE-9. Here we provide additional examples from our papers and on-going work.

The first figure of this section, taken from DeGennaro et al. (2009), shows Hyades CMDs with three sets of stellar evolution models placed at their average fit values as determined by a previous version of the code, BASE-8. Because these stellar models do not provide good fits to the lower main sequence, the following figure shows the derived age from BASE-8 for each of the three input models and for a range of lower main sequence cut-offs. In this way DeGennaro et al. were able to argue that their derived parameters were stable over an appropriate range of data and were able to quantitatively point to where problems emerged in the models.

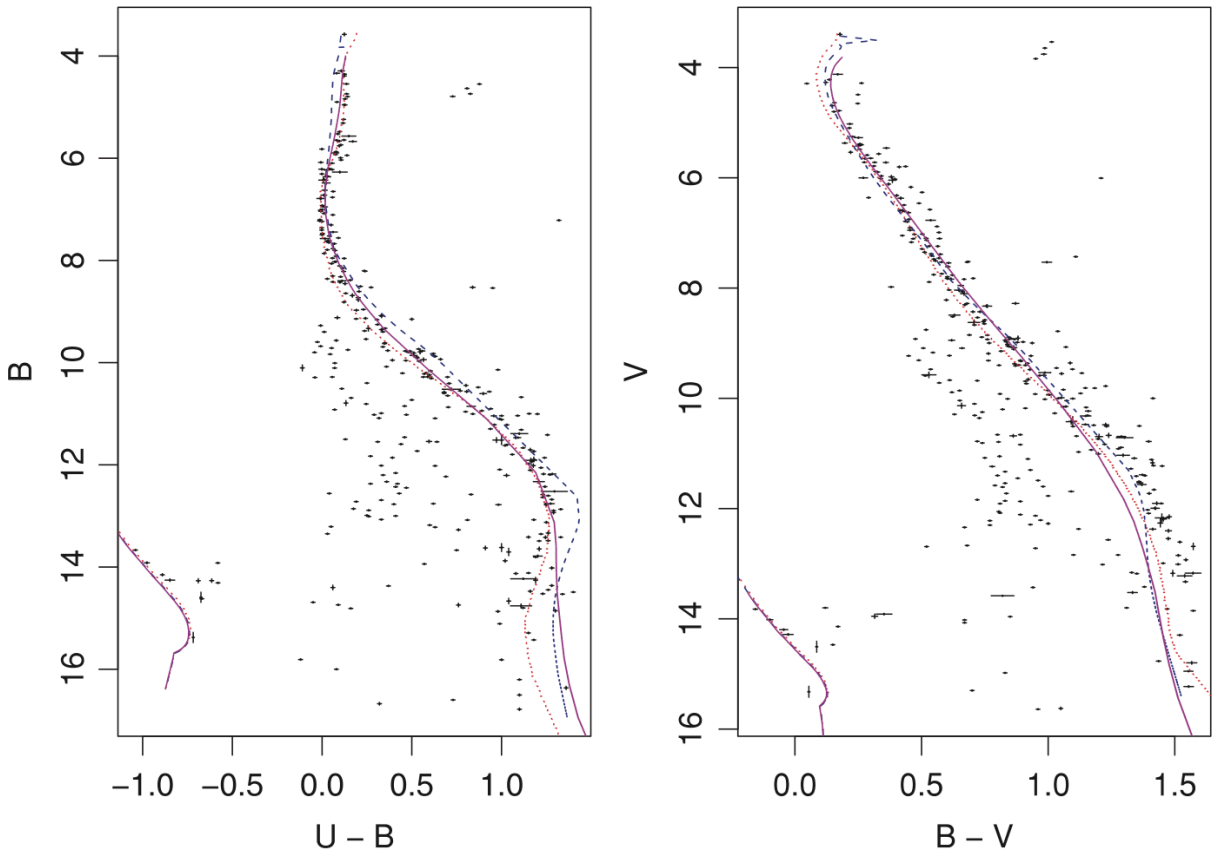

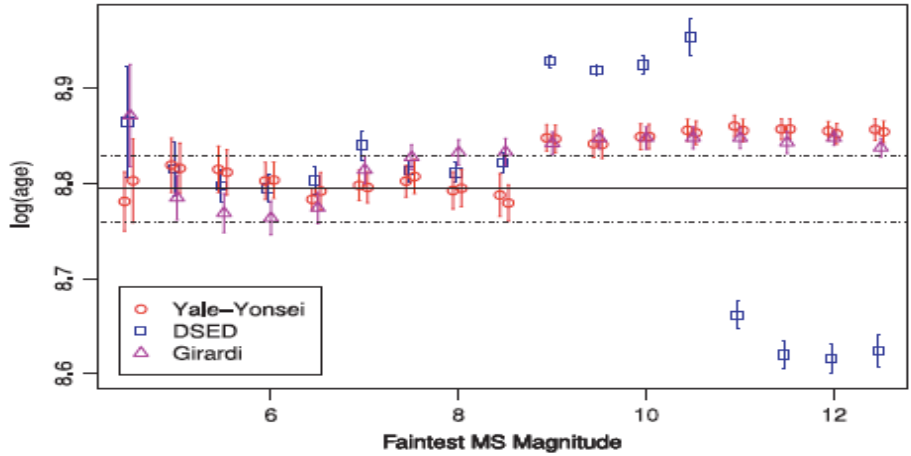

This next figure, taken from Jeffery, E. J. (2009, Ph.D. Dissertation, University of Texas at Austin) compare the age information resident in just the main sequence turn-off stars (black dashed line) compared to that resident in the white dwarfs. Data from the main sequence was included in both BASE-8 runs, and this provides the primary constraints on metallicity, distance, and reddening. This is useful for studying the information content in the MSTO vs. WD regions.

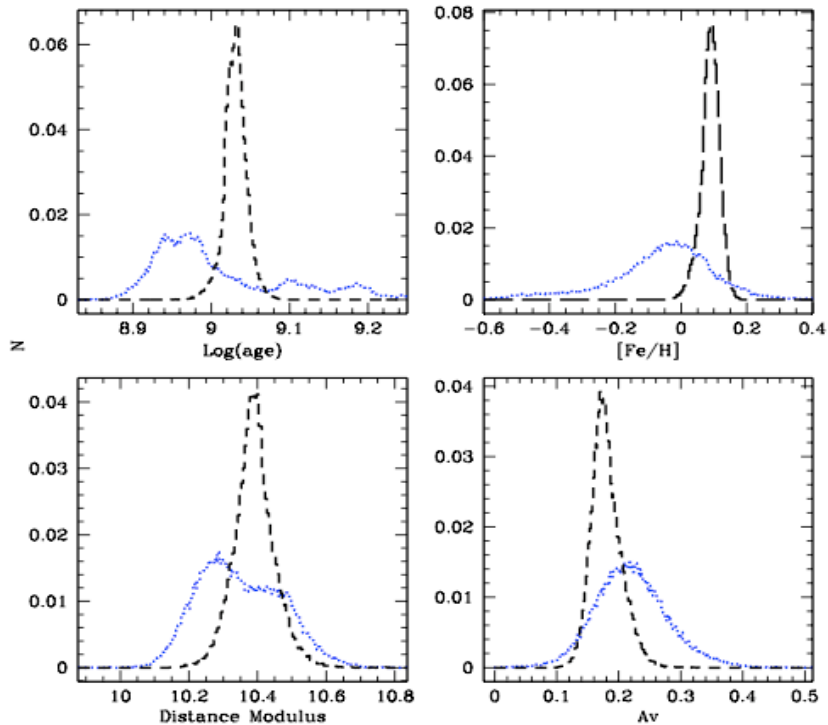

Figure 5.16: A comparison of MCMC results for cluster parameters of NGC 2360 from fitting the MSTO vs. WDs. The dashed black lines are the posterior distributions from fitting the MSTO while the blue dotted lines are from fitting the WDs.

The next figure, also from Jeffery (2009), indicates how one can study the sensitivity of a given result to the observations of an individual star. For the open cluster NGC 2360, the posterior age distribution is given by the black line. During some iterations, however, a particular WD is fit as a field star and the remaining WDs yield the posterior age distribution indicated in red. During the iterations when this particular WD is included in the fit, the posterior age distribution is as indicated in blue. The final age posterior distribution is a linear combination of these two distributions based on the fraction of time this particular WD was included in the fit.

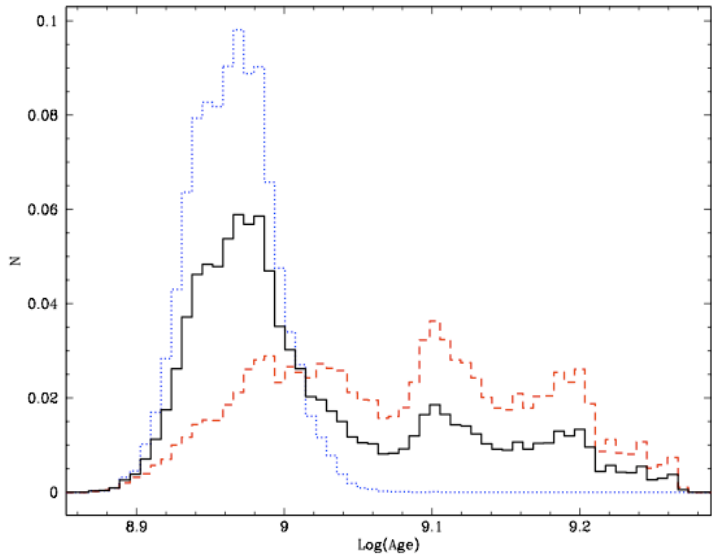

Figure 5.7: Age distribution of NGC 2360 with and without the inclusion of WD5. The solid black line is the complete age posterior distribution. The dotted blue line is the age distribution when WD5 is included as a cluster member. The red dashed line is the distribution when WD5 is excluded as a field star. This indicates the importance of this star in determining the location of the WD cooling sequence and hence measuring the age via MCMC.

The next figure shows unpublished work based on applying BASE-8 to an individual WD. In this particular case, we know that the WD has a hydrogen atmosphere (type DA), yet for demonstration purposes we analyze it both as a DA and as a DB (helium atmosphere). We also try two different initial-final mass relations (from Salaris et al. 2009 and Williams, Bolte, & Koester 2009). The clouds of points show acceptable fits and the error bars indicate the mean and standard deviation for each of the four cases. Clearly these distributions are non-Gaussian and publishing just the means and standard deviations could lead readers to misunderstand the results. This kind of analysis can also point the way toward future observational work. For this star a trigonometric parallax could potentially rule out much of the age range, yielding a precise age. If this star were a DB a much more accurate trig parallax would be required to meaningfully constrain the age. This is not a general statement about WDs, but a result for this star with the available photometry (grizJHK).

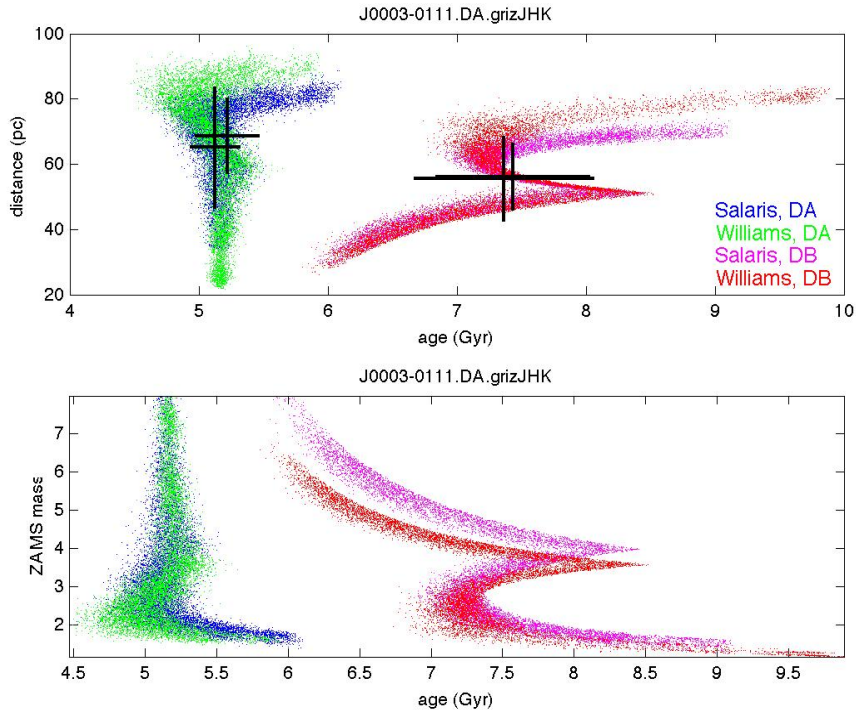

## VII. Modifying the code to extend its capabilities

We continue to upgrade BASE-9 for our on-going projects. If you wish to add capability to BASE-9, we will be happy to suggest to you how best to go about this and try to estimate the work involved. Here is an example list of how involved a variety of tasks are likely to be.

*Less than 2 hours*: Modifying the IFMR. You can do this by editing or adding a few lines of code in `ifmr.cpp`.

*Less than 8 hours:* Change the IMF. You will need to create a subroutine where a random mass value can be drawn from your IMF distribution. This currently takes place in `drawFromIMF.cpp`. Note that you will also have to normalize the IMF for the Bayesian routine to work properly and that this takes place in `densities.cpp` and is stored in `logMassNorm`.

*Less than 16 hours:* Incorporating another set of stellar evolution models – see instructions at the top of `msRgbEvol.cpp` and possibly `wdCooling.cpp` and/or `gBergMag.cpp`.

*Less than a week:* Sampling a new variable (e.g. stellar rotation, alpha-element enhancement). This takes place primarily in `singlePopMcmc/MpiMcmcApplication.cpp` and `base9/densities.cpp`.